\documentclass[referee]{raa}           
\usepackage{graphicx,times}
\usepackage{natbib}
\usepackage{amssymb,amsmath}
\usepackage{longtable}
\usepackage{geometry}
\bibpunct{(}{)}{;}{a}{}{,}

\usepackage[a4paper=true,dvipdfm=true,pagebackref=true]{hyperref}
\hypersetup{pdftitle = The title of my PDF, pdfauthor = My name, pdfsubject= The subject, pdfkeywords = keyword1 keyword2 keyword3} 
\hypersetup{colorlinks = true, linkcolor = green, anchorcolor = red, citecolor = blue, filecolor = red, pagecolor = red, urlcolor = red}

\begin{document}

   \title{Observations and light curve solutions of the W~UMa binaries\\
    V796 Cep, V797 Cep, CSS J015341.9+381641 and NSVS 3853195
}

 \volnopage{ {\bf 2012} Vol.\ {\bf X} No. {\bf XX}, 000--000}
   \setcounter{page}{1}

 \author{D. P. Kjurkchieva\inst{1}, V. A. Popov\inst{1,2}, S. I. Ibryamov\inst{1}, D. L. Vasileva\inst{1}, N. I. Petrov\inst{3}
   }
%% Here is an example of three authors come from different institutes.
%% For single author or all the authors from an institute, use "\inst{}" only

   \institute{ Department of Physics, Shumen University, 115 Universitetska Str., 9712 Shumen, Bulgaria;
{\it d.kjurkchieva@shu.bg}\\
%% Please give the E-mail address of the author, to whom future correspondence and
%% offprint requests will be sent.
        \and
             IRIDA Observatory, Rozhen NAO, Bulgaria\\
    \and
%     Center for Astrophysics, University of Science and Technology of China, Hefei 230026, China\\
Institute of Astronomy and NAO, Bulgarian Academy of Sciences, 72 Tsarigradsko Shose Blvd., 1784 Sofia, Bulgaria\\
\vs \no
   {\small Received ; Аccepted }
}

\abstract{Photometric observations in Sloan \emph{g'} and
\emph{i'} bands of four W UMa binaries, V796 Cep, V797 Cep, CSS
J015341.9+381641 and NSVS 3853195, are presented. Our observations
showed that CSS J015404.1+382805 and NSVS 3853195 are the same
star. We determined the initial epochs $T_0$ of all targets and
improved the period of NSVS 3853195. The light curve solutions of
our data revealed that the components of each target are almost
the same in mass, temperature, radius and luminosity. The stellar
components are of G and K spectral types and undergo partial
eclipses. All systems have barely-overcontact configurations and
belong to H subtype W UMa binaries. We established that the
relation between the luminosity ratio $l_2/l_1$ and mass ratio
\emph{q} of our targets is approximately $l_2/l_1 = q^{1.5}$.
 }

   \authorrunning{Kjurkchieva et al.}            %author_head in even pages
   \titlerunning{Light curve solutions of four W UMa binaries}  % title_head in odd pages
   \maketitle

%________________________________________________ sections below
%
\section{Introduction}           %% first-level sections will be auto-capitalized
\label{sect:intro}

The creation of stellar evolutional scheme requires a knowledge of
the fundamental parameters of stars in different stages of their
evolution. Eclipsing binary systems, especially W UMa binaries,
are the most important sources of such information. It is supposed
that they are result from the evolution of wide binaries by
angular momentum loss and mass-ratio reversals
(\citealt{Stepien+2006}, \citealt{Qian+2003}). Around 25 $\%$ of
main-sequence star binaries have separations small enough so when
their primaries ascend the giant branch, mass transfer via
Roche-lobe overflow puts the begining of a common envelope phase
(\citealt{Willems+Kolb+2004}). At this stage the two stars orbit
within a single envelope of material, quickly losing angular
momentum and spiraling towards each other (\citealt{Webbink+1984};
\citealt{Ivanova+etal+2013}). The common-envelope phase is
probably a short-lived stage that ends by envelope ejection and a
tighter binary or by a merger. But understanding of common
envelope stage remains one of the most important unsolved problems
in stellar evolution (\citealt{Ivanova+etal+2013}).

The components in a W UMa system have nearly equal surface
temperatures in spite of their often greatly different masses
(\citealt{Binnendijk+1965}). The model of \cite{Lucy+1968a},
\cite{Lucy+1968b} explained this effect by a common convective
photosphere which embedded two main sequence stars. As a result
one should expect the observable luminosities to have another
dependence on the mass ratio than would be the case of two main
sequence stars in detached configuration. The condition for equal
surface temperatures leads to specific period-luminosity-color
(PLC) relations of W UMa stars (\citealt{Rucinski+1994},
\citealt{Rucinski+Duerbeck+1997}) allowing currently to predict
their absolute magnitudes $M_V$ to about 0.25 mag
(\citealt{Rucinski+2004}). The PLC relations combined with the
ease detection make these binaries useful tracers of distance
(\citealt{Klagyivik+Csizmadia+2004}; \citealt{Gettel+etal+2006};
\citealt{Eker+etal+2009}). Moreover, the W UMa stars are important
targets for the modern astrophysics because they give information
for the processes of tidal interactions, mass loss and mass
transfer, angular momentum loss, merging or fusion of the stars
(\citealt{Martin+etal+2011}).

In this paper we present photometric observations and light curve
solutions of four W UMa binaries: V796 Cep (GSC 04502-00138,
TYC 4502-138-1), V797 Cep (GSC 04502-01040, 2MASS
J01424764+8007522), CSS J015341.9+381641, NSVS 3853195
(CSS J015404.1+382805). Table~1 presents the coordinates of our
targets and available information for their light variability.

\begin{table}
\begin{minipage}[t]{\textwidth}
\caption{Parameters of our targets from the VSX database}
\label{Tab1}
\centering
\begin{scriptsize}
 \begin{tabular}{clcllcc}

  \hline\hline
Target               & RA          & Dec         & Period    & V     & Amplitude  & Type     \\
                     &             &             & [d]       & [mag] & [mag] &           \\
  \hline
  \noalign{\smallskip}
V796 Cep             & 01 41 36.39 & +80 04 19.1 & 0.3929661  & 12.20 & 0.65 &  EW  \\
V797 Cep             & 01 42 47.64 & +80 07 52.3 & 0.270416   & 14.60 & 0.40 &  EW \\
CSS J015341.9+381641 & 01 53 41.95 & +38 16 41.1 & 0.347518   & 13.47 & 0.40 &  EW   \\
NSVS 3853195         & 01 54 04.05 & +38 28 05.2 & 0.29253    & 13.52 & 0.39 &  EW  \\
\hline\hline
  \end{tabular}
\end{scriptsize}
\end{minipage}
\end{table}

\begin{table}
\begin{minipage}[t]{\textwidth}
\caption{Journal of our photometric observations}
\label{Tab2}
\centering
\begin{scriptsize}
\begin{tabular}{ccrrrr}
\hline\hline
Target&  Date        & Exposure $g'$ & Exposure $i'$ & Number $g'$ & Number $i'$ \\
\hline
V796 Cep, V797 Cep   & 2015 Oct 25 & 90& 120 & 125& 125 \\
                     & 2015 Oct 26 & 90& 120 &  84&  82 \\
                     & 2015 Oct 27 & 90& 120 &  60&  59 \\
                     & 2015 Oct 28 & 90& 120 & 146& 146 \\
CSS J015341.9+381641, NSVS 3853195 & 2015 Nov 07 & 60 & 90 & 85 &  84 \\
                     & 2015 Nov 08 & 60&  90 &  67&  75 \\
                     & 2015 Nov 11 & 60&  90 &  84&  84 \\
                     & 2015 Nov 12 & 60&  90 &  43&  42 \\
                     & 2015 Nov 13 & 60&  90 &  84&  85 \\
 \hline\hline
\end{tabular}
\end{scriptsize}
\end{minipage}
\end{table}

\begin{table}
\begin{minipage}[t]{\textwidth}
\caption{List of the standard stars}
\label{Tab3}
\centering
\begin{scriptsize}
\begin{tabular}{ccccll}
\hline\hline
 Label & Star ID & RA & Dec & \emph{g'} & \emph{i' }\\
\hline
Target 1 & V0796 Cep & 01 41 36.39 & +80 04 19.10 & 12.320  & 11.789  \\
Target 2 & V0797 Cep & 01 42 47.64 & +80 07 52.30 & 14.966  & 14.025  \\
Chk & UCAC4 851-002007 & 01 41 16.52 & +80 04 21.76 & 13.755 (0.010) & 13.018 (0.010) \\
C1 & UCAC4 851-002085 & 01 45 07.01 & +80 10 45.03 & 13.238 (0.011) & 12.534 (0.011) \\
C2 & UCAC4 851-002011 & 01 41 28.03 & +80 11 18.42 & 13.870 (0.010) & 13.351 (0.012) \\
C3 & UCAC4 851-002002 & 01 40 56.97 & +80 04 14.51 & 13.448 (0.009) & 12.844 (0.010) \\
C4 & UCAC4 851-002062 & 01 43 56.73 & +80 02 08.49 & 14.205 (0.011) & 13.452 (0.013) \\
C5 & UCAC4 850-002063 & 01 41 51.38 & +79 56 58.55 & 13.257 (0.007) & 12.651 (0.009) \\
C6 & UCAC4 851-002028 & 01 42 25.41 & +80 01 00.68 & 13.898 (0.009) & 13.058 (0.009) \\
\hline
Target 3 & CSS J015341.9+381641 & 01 53 41.95 & +38 16 41.10 & 13.897 & 13.117 \\
Target 4 & NSVS 3853195 & 01 54 04.05 & +38 28 05.26 & 14.054 & 13.178 \\
Chk & UCAC4 643-007188 & 01 54 30.68 & +38 29 00.15 & 13.104 (0.014) & 11.754 (0.009) \\
C1 & UCAC4 644-007104 & 01 54 12.54 & +38 36 49.06 & 13.975 (0.016) & 13.902 (0.018) \\
C2 & UCAC4 643-007165 & 01 54 04.61 & +38 35 54.88 & 14.112 (0.011) & 13.419 (0.015) \\
C3 & UCAC4 643-007180 & 01 54 23.95 & +38 35 42.60 & 14.061 (0.014) & 13.369 (0.015) \\
C4 & UCAC4 643-007204 & 01 54 50.23 & +38 33 52.57 & 13.971 (0.019) & 13.147 (0.015) \\
C5 & UCAC4 643-007182 & 01 54 26.03 & +38 29 38.62 & 13.720 (0.009) & 13.012 (0.011) \\
C6 & UCAC4 643-007126 & 01 53 26.47 & +38 30 02.18 & 13.702 (0.016) & 12.978 (0.013) \\
C7 & UCAC4 642-006881 & 01 54 12.52 & +38 23 56.10 & 13.937 (0.012) & 12.907 (0.011) \\
C8 & UCAC4 642-006842 & 01 53 30.38 & +38 20 34.86 & 13.929 (0.024) & 13.071 (0.018) \\
C9 & UCAC4 642-006908 & 01 54 36.06 & +38 20 04.67 & 14.068 (0.013) & 11.491 (0.014) \\
C10 & UCAC4 642-006921 & 01 54 46.30 & +38 19 35.13 & 13.161 (0.021) & 11.764 (0.014) \\
C11 & UCAC4 643-007147 & 01 53 51.89 & +38 28 10.11 & 12.849 (0.018) & 12.191 (0.014) \\
 \hline
\end{tabular}
\end{scriptsize}
\end{minipage}
\end{table}

\section{Observations}

Our CCD photometric observations of the targets in Sloan \emph{g',
i'} bands were carried out at Rozhen National Astronomical
Observatory with the 30-cm Ritchey-Chr\'{e}tien Astrograph
(located into the \emph{IRIDA South} dome) using CCD camera ATIK
4000M (2048 $\times$ 2048 pixels, 7.4 $\mu$m/pixel, field of view
35 $\times$ 35 arcmin). Information for our observations is
presented in Table~2. In fact, the pairs V796 Cep, V797 Cep and
CSS J015341.9+381641, NSVS 3853195 fall on two observed fields
(see coordinates in Table 1).

The data were obtained during photometric nights with seeing
within 1.1--1.9 arcsec and humidity below 70 $\%$. Twilight flat
fields were obtained for each filter, dark and bias frames were
also taken throughout the run. The frames were combined
respectively into a single master bias, dark and flat frames. The
standard procedure was used for reduction of the photometric
data (de-biasing, dark frame subtraction and flat-fielding) by
software \textsc{AIP4WIN2.0} (\citealt{Berry+Burnell+2006}).

We used aperture photometry with radius of 1.5 FWHM of the star
image, along with sky background measurements with annuli
enclosing a comparable area. The light variability of the targets
was estimated with respect to nearby comparison (constant) stars
in the observed field of each target, so called ensemble
photometry. A check star served to determine the observational
accuracy and to check constancy of the comparison stars. The CCD
ensemble photometry calculates the difference between the
instrumental magnitude of the target and a comparison magnitude
obtained from the mean of the intensities of the comparison stars.
The use of numerous comparison stars scattered over the CCD field
increases considerably the statistical accuracy of the comparison
magnitude (\citealt{Gilliland+Brown+1988},
\citealt{Honeycutt+1992}).

We performed the ensemble aperture photometry with the software
\textsc{VPHOT}. Table~3 presents the coordinates of the comparison
and check stars of our targets from the catalogue UCAC4
(\citealt{Zacharias+etal+2013}) and their magnitudes from the
catalogue APASS DR9 (\citealt{Henden+2016}). The values in
brackets correspond to the standard deviations of the standard
stars during the observational nights. The choice of comparison
and check stars in the same field of view of the targets means
practically equal extinctions for all stars. The transformation of
the obtained instrumental magnitudes to standard ones was made
manually. For this aim we used the mean color of the ensemble
comparison-star $\overline{(g'-i')}_{comp}$ and transformation
coefficients of our equipment (calculated earlier using standard
star field M67). The calculated corrections of the instrumental
magnitudes for our targets were from $-$0.0008 mag to 0.0003 mag
in $g'$ filter (within the observational precision) and from
$-$0.0258 mag to 0.0085 mag in $i'$ filter.

%Their values applicable to the presented observations are: $T_{g', g'i'} = -0.002\pm0.012$,
%$T_{i', g'i'} = -0.061\pm0.017$, $T_{g'i'} = 1.063 \pm 0.011$. They show that our local photometric system is very close to the
%standard Sloan system (especially $g'$).

\begin{figure}
   \centering
   \includegraphics[width=8cm, angle=0]{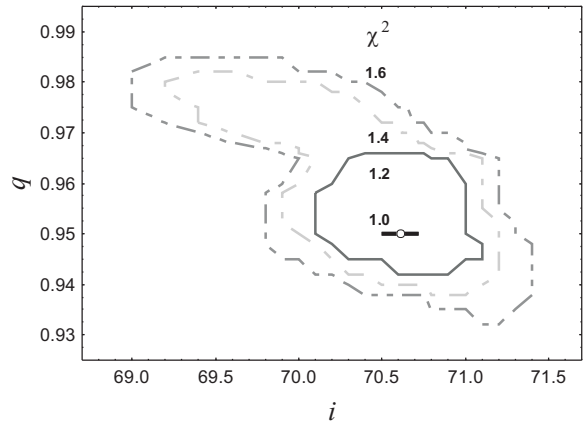}
  % \begin{minipage}[]{85mm}
   \caption{Illustration of the \emph{q}-search analysis for V796 Cep:
   the different isolines circumscribe the areas whose
normalized $\chi^2$ are smaller than the marked values; the empty
circle corresponds to the final value of the mass ratio and
orbital inclination given in Table 5.}
%\end{minipage}
   \label{Fig1}
   \end{figure}

\begin{figure}
\centering
\includegraphics[width=14cm,scale=1.00]{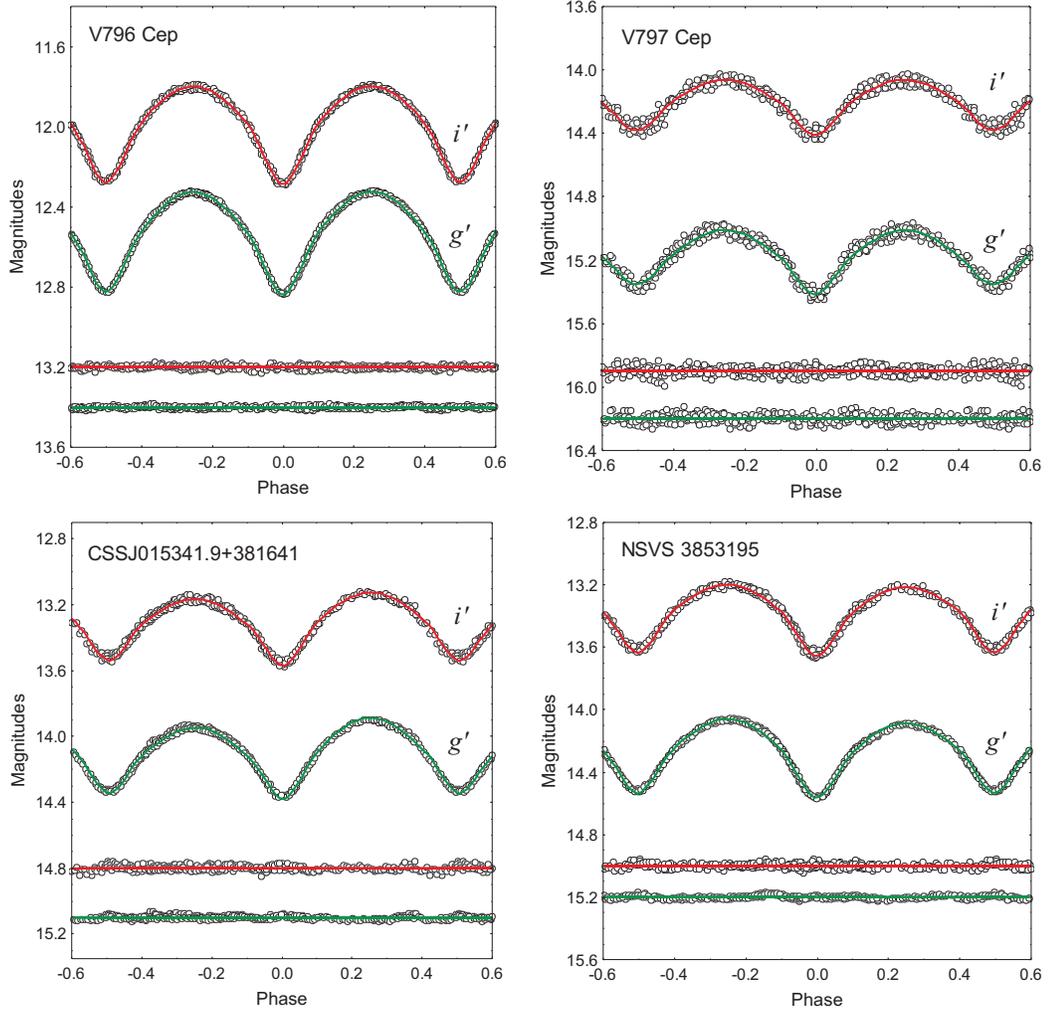}
\caption{The folded light curves of the targets with their fits
and the corresponding residuals (shifted vertically by different amount to save space).}
\label{Fig2}
\end{figure}

\begin{figure}
\begin{center}
\includegraphics[width=14cm,scale=1.00]{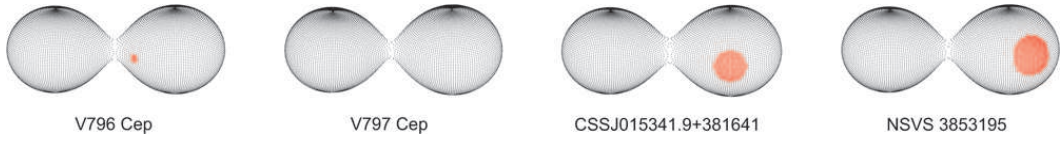}
\caption[]{3D configurations of the targets}
\label{Fig3}
\end{center}
\end{figure}

We determined the times of the individual minima (Table 4)
by the method of \cite{Kwee+Woerden+1956}.

\begin{table}
\begin{center}
\caption[]{Times of minima of our targets \label{Tab4}}
\begin{scriptsize}
\begin{tabular}{cccc}
\hline\hline \noalign{\smallskip}
Target & Min. I & Min. II & IRIDA cycle \\
\hline
V0796 Cep   & 2457321.43582(9) & - &  0.0 \\
        & - & 22457322.41776(13) &  2.5 \\
        & 2457323.40045(19) & - &  5.0 \\
        & - & 2457324.38263(8) &  7.5 \\
        & 2457324.57929(1) & - &  8.0 \\
\hline
V0797 Cep   & 2457321.31715(74) & - &  0.0 \\
        & - & 2457321.45201(25) &  0.5 \\
        & 2457321.58629(22) & - &  1.0 \\
        & - & 2457322.26135(33) &  3.5 \\
        & 2457322.39815(24) & - &  4.0 \\
        & - & 2457323.34511(22) &  7.5 \\
        & 2457324.29034(18) & - &  11.0 \\
        & - & 2457324.42626(3) &  11.5 \\
\hline
CSS J015341.9+381641    & - & 2457333.44320(11) &  -0.5 \\
        & 2457333.61496(31) & - &  0.0 \\
        & - & 2457334.48599(14) &  2.5 \\
        & 2457338.47982(18) & - &  14.0 \\
        & - & 2457340.39396(31) &  19.5 \\
        & 2457340.56578(15) & - &  20.0 \\
\hline
NSVS 3853195 & - & 2457333.44266(9) &  -0.5 \\
        & 2457333.59001(21) & - &  0.0 \\
        & 2457334.46774(15) & - &  3.0 \\
        & - & 2457338.41610(11) &  16.5 \\
        & 2457338.56296(13) & - &  17.0 \\
        & 2457339.44062(12) & - &  20.0 \\
        & - & 2457340.46366(14) &  23.5 \\
 \hline\hline
\end{tabular}
\end{scriptsize}
\end{center}
\end{table}

\section{Light curve solutions}

The light curves of our targets were solved by the code
\textsc{PHOEBE} (\citealt{Prsa+Zwitter+2005}). It is based on the
Wilson--Devinney (WD) code (\citealt{Wilson+Devinney+1971},
\citealt{Wilson+1979}) but also provides a graphical user
interface and modeling in Sloan filters of our observations. We
used the traditional convention the Min. I (phase 0.0) to be the
deeper light minimum and the star that is eclipsed at Min. I to be
a primary component.

Mean temperatures $T_{m}$ of the binaries were determined in
advance (see Table 6 further) on the basis of their infrared color
indices \emph{(J-K)} from the 2MASS catalog and the calibration
color-temperature of \cite{Tokunaga+2000}. The preliminary runs
revealed that all targets are overcontact systems. Hence, we
applied mode ''Overcontact binary not in thermal contact'' of the
code. Firstly we fixed $T_{1}$ = $T_{m}$ and varied the initial
epoch $T_{0}$ and period \emph{P} to search for fitting the phases
of light minima and maxima. After that we fixed their values and
varied simultaneously secondary temperature $T_{2}$, orbital
inclination $i$, mass ratio $q$ and potential $\Omega$ to search
for reproducing of the whole light curves. The data in \emph{i'}
and \emph{g'} bands were modelled simultaneously.

We adopted coefficients of gravity brightening $g_1=g_2$ = 0.32
and reflection effect $A_1=A_2$ = 0.5 appropriate for late-type
stars while the linear limb-darkening coefficients for each
component and each color were updated according to the tables of
\cite{vanHamme+1993}. Solar metallicity was assumed for the
targets because they consist of late stars from the solar
vicinity. In order to reproduce the light curve distortions we
used cool spots whose parameters (longitude $\lambda$, angular
size $\alpha$ and temperature factor $\kappa$) were adjusted.

After reaching the best solution we varied together all parameters
($T_{2}$, $i$, $q$, $\Omega$, $T_{0}$ and \emph{P}) around the
values from the last run and obtained the final model. In order to
determine stellar temperatures $T_{1}$ and $T_{2}$ around the mean
value T$_{m}$ we used the formulae
(\citealt{Kjurkchieva+etal+2016a}):
\begin{equation}
T_{1}=T_m + \frac{\Delta T}{c+1},
\end{equation}
\begin{equation}
T_{2}=T_1 -\Delta T,
\end{equation}
where $c=l_2/l_1$ (luminosity ratio) and $\Delta T=T_m-T_2^{PH}$
were taken from the last \textsc{PHOEBE} fitting.

Although \textsc{PHOEBE} (as WD) works with potentials, it gives a
possibility to calculate all values (polar, point, side, and back)
of relative radius $r_i=R_i/a$ of each component ($R_i$ is linear
radius and \emph{a} is orbital separation). In the absence of
radial velocity curves we put as default \emph{a} = 1 because from
photometry only we cannot determine binary separation. Moreover,
\textsc{PHOEBE} yields as output parameters bolometric magnitudes
$M_{bol}^i$ of the two components in conditional units (when
radial velocity data are not available). But their difference
$M_{bol}^2-M_{bol}^1$ determines the true luminosity ratio
$c=L2/L1=l2/l1$. Fillout factor $f = [\Omega -
\Omega(L_1)]/[\Omega(L_2) - \Omega(L_1)]$ can be also calculated
from the output parameters of PHOEBE solution.

In order to take into account the effect of expected correlation
between the mass ratio and orbital inclination we carried out
\emph{q}-search analysis as described in
\cite{Kjurkchieva+etal+2016a}. For this aim we fixed the component
temperatures and radii as well as the spot parameters and
calculated the normalized $\chi^{2}$ for a two-dimensional grid
along $i$ and $q$. Figure 1 illustrates the result from this
\emph{q}-search procedure for the target V796 Cep.

Table~5 contains final values of the fitted stellar parameters and
their \textsc{PHOEBE} uncertainties: initial epoch $T_{0}$; period
\emph{P}; mass ratio \emph{q}; inclination \emph{i}; potential
$\Omega$; secondary temperature $T_{2}^{PH}$. Table 6 exhibits the
calculated parameters: stellar temperatures $T_{1, 2}$; stellar
radii $r_{1, 2}$ (back values); fillout factor \emph{f}; ratio of
relative stellar luminosities $l_2/l_1$. Their errors are
determined from the uncertainties of output parameters used for
their calculation. Table~7 gives information for the spot
parameters. The synthetic light curves corresponding to our
solutions are shown in Fig.~2 as continuous lines.

\begin{table}
\centering
\caption[]{Values of the fitted parameters}\vspace{0.1in}
\label{Tab5}
  \begin{scriptsize}
 \begin{tabular}{cllcccc}
\hline
Star                & $T_0$              &  \emph{P}   & \emph{q} &\emph{i} &$\Omega$ & $T_2^{PH}$   \\
\hline
V796 Cep            &  2457321.43582(9)  & 0.392966    & 0.948(2) & 70.7(1) &  3.612(7) & 6400(19) \\
V797 Cep            &  2457321.31715(74) & 0.270416    & 0.886(2) & 64.7(1) &  3.525(2) & 4625(42) \\
CSS J015341.9+381641&  2457333.61496(31) & 0.347518    & 0.892(2) & 70.0(2) &  3.490(1) & 5607(28) \\
NSVS 3853195        &  2457333.59001(21) & 0.292524(4) & 0.899(2) & 69.8(1) &  3.539(3) & 5592(30) \\
\hline
\end{tabular}
\end{scriptsize}
\end{table}

\begin{table}
\centering
\caption{Calculated parameters }\vspace{0.1in}
\label{Tab5}
\begin{scriptsize}
 \begin{tabular}{cccccccc}
 \hline
Target              & $T_m$&  $T_1$     &   $T_2$      &   $r_1$  &   $r_2$  & \emph{f} &   $l_2/l_1$   \\
\hline
V796 Cep            & 6407 & 6410(19)   &   6403(19)   &   0.421(1)  &   0.412(1)  &  0.101  &   0.951  \\
V797 Cep            & 4770 & 4833(44)   &   4688(42)   &   0.424(1)  &   0.403(1)  &  0.075  &   0.771  \\
CSS J015341.9+381641& 5715 & 5765(29)   &   5657(28)   &   0.434(1)  &   0.414(1)  &  0.166  &   0.867  \\
NSVS 3853195        & 5688 & 5733(31)   &   5637(30)   &   0.425(1)  &   0.406(1)  &  0.089  &   0.865  \\
\hline
\end{tabular}
\end{scriptsize}
\end{table}

\begin{table}
\begin{minipage}[t]{\columnwidth}
\caption{Parameters of the cool spots of the targets} \label{tab6}
\centering
 \begin{scriptsize}
 \begin{tabular}{ccrrc}
\hline\hline
Star                & $\beta$   &   $\lambda$   &   $\alpha$ &  \emph{k }  \\
\hline
V796 Cep            &   90(5)  &   35(1)  &   5.0(1)   &   0.90(1) \\
CSS J015341.9+381641&   90(5)  &   90(1)  &   20.0(1)  &   0.80(1) \\
NSVS 3853195        &   80(5)  &   120(1) &   25.0(2)  &   0.95(1) \\
\hline\hline
\end{tabular}
\end{scriptsize}
\end{minipage}
\end{table}

The mean (\emph{g', i'}) residuals for the final fittings are:
(0.005, 0.007) for V796 Cep; (0.021, 0.022) for V797 Cep; (0.009,
0.012) for CSS J015341.9+381641; (0.009, 0.013) for NSVS 3853195.
The mean (\emph{g', i'}) residuals of the standard stars (Table 3)
for the first and second pairs of targets are correspondingly
(0.010, 0.011) and (0.017, 0.015). Hence, our fittings are
excellent for the three targets and very good for the faint star
V797 Cep (Fig. 2). The small imperfectness of our modeling may due
to inadequate treatment of the overcontact binaries
(\citealt{Prsa+etal+2016}) and to long exposures
(\citealt{Kipping+2010}).

\begin{figure}
\begin{center}
\includegraphics[width=6cm,scale=1.00]{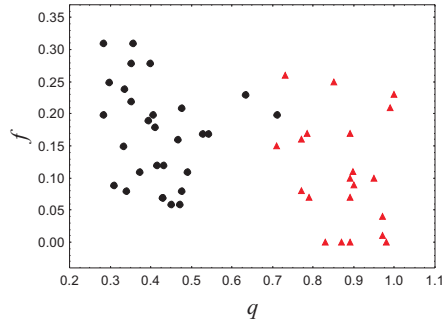}
\caption[]{Distribution fillout factor-- mass ratio of W UMa
stars: red triangles are for shallow contact high mass ratio
targets; black circles are for targets with decreasing periods
from the sample of Yang (2013)} \label{Fig4}
\end{center}
\end{figure}

\section{Conclusions}

The main results from the light curve solutions of our data are as
follows.

(1) We determined the initial epochs $T_0$ of the four targets
(Table 5).

(2) We improved the period of NSVS 3853195 (Table 5) on the base
of all photometric data: CRTS, NSVS, SWASP and IRIDA. The previous
period values of the other three targets fitted well our data.

(3) Our observations revealed that CSS~J015404.1+382805 and NSVS
3853195 are the same star (while VSX identified two stars).

(4) The components of each target are almost the same in mass,
temperature, radius and luminosity (Tables 5-6).

(5) The stellar components of all targets are of G and K spectral
types and they undergo partial eclipses.

(6) All targets have overcontact configurations with small fillout
factor (Fig. 3, Table 6). This means that they probably
are newly formed contact binaries (\citealt{Qian+etal+2014}).

(7) Three binaries revealed O'Connell effect that was reproduced
by cool spots (Table~7) on their primary components. They are
appearances of the magnetic activity of these targets.

%CSTAR 038663 (1/q=0.89; f=10.6%; 0.27 4500 Qian, S.-B. et al.; 2014ApJS..212....4  l2/l1 1.13=0.89^x  x=

%BI Vul (q=1; f=4.4%; 0.25 d, 4500  Qian, S.-B.;Qian, S.-B.et al; 2013ApJS..209...13)

% AD Cnc (q=0.77; f=8.3%; 4900 K 0,28 d Qian, S.-B.et al; 2007ApJ...671..811).

(8) All our targets have mass ratio $q \geq$ 0.88 (Table~5), i.e.
they belong to the H subtype W UMa systems (with $q \geq$ 0.72).
\cite{Csizmadia+Klagyivik+2004} revealed that the different
subtypes of W UMa's are located into different regions on the mass
ratio -- luminosity ratio diagram (their fig. 1) but above the
line $l_2/l_1 = q^{4.6}$ representing the mass-luminosity relation
for MS detached stars. Our targets support this conclusion and the
relation between their mass ratio and luminosity ratio is $l_2/l_1 = q^{1.5}$, i.e. close
to that of \cite{Lucy+1968a}.

\begin{table}
\begin{minipage}[t]{\columnwidth}
\caption{Parameters of the cool spots of the targets} \label{tab6}
\centering
 \begin{scriptsize}
 \begin{tabular}{ccrcc}
\hline\hline
Star                        & $q$ & $l_2/l_1$ & $f$ & References   \\
\hline

AD Cnc                      & 0.77 &  1.00   & 0.08 & \cite{Qian+etal+2007}   \\
BI Vul                      & 0.97 &  1.22   & 0.04 & \cite{Qian+etal+2013}   \\
CSTAR 038663                & 0.89 &  1.13   & 0.10 & \cite{Qian+etal+2014}   \\

1SWASP J174310.98+432709.6  & 1.00 &  0.65   & 0.23 & \cite{Kjurkchieva+etal+2015a}\\
NSVS 11234970               & 0.99 &  0.55   & 0.21 & \cite{Kjurkchieva+etal+2015a}\\
NSVS 11504202               & 0.98 &  0.71   & 0.00 & \cite{Kjurkchieva+etal+2015a}\\
NSVS 11534299               & 0.87 &  0.77   & 0.00 & \cite{Kjurkchieva+etal+2015a}\\

NSVS 1776195                & 0.83 &  0.96   & 0.00 & \cite{Kjurkchieva+etal+2015b}\\
NSVS 113026                 & 0.79 &  1.00   & 0.07 & \cite{Kjurkchieva+etal+2015b}\\

NSVS 2244206                & 0.73 &  0.53   & 0.26 & \cite{Kjurkchieva+etal+2016b}\\
NSVS 908513                 & 0.71 &  0.60   & 0.15 & \cite{Kjurkchieva+etal+2016b}\\
VSX J062624.4+570907        & 0.77 &  0.63   & 0.16 & \cite{Kjurkchieva+etal+2016b}\\

CSS J171508.5+350658        & 0.89 &  0.64   & 0.00 & \cite{Kjurkchieva+etal+2016a}\\

USNO-B1.0-1395-0370184      & 0.97 &  0.90   & 0.01 & \cite{Kjurkchieva+etal+2016c}\\
USNO-B1.0-1395-0370731      & 0.85 &  0.83   & 0.25 & \cite{Kjurkchieva+etal+2016c}\\

NSVS 2459652                & 0.786&  0.73   & 0.17 & \cite{Kjurkchieva+etal+2016d}\\
NSVS 7377875                & 0.898&  0.84   & 0.11 & \cite{Kjurkchieva+etal+2016d}\\

V796 Cep                    & 0.95 &  0.95   & 0.10 & this paper   \\
V797 Cep                    & 0.89 &  0.77   & 0.07 & this paper   \\
CSS J015341.9+381641        & 0.89 &  0.87   & 0.17 & this paper   \\
NSVS 3853195                & 0.90 &  0.86   & 0.09 & this paper   \\

\hline\hline
\end{tabular}
\end{scriptsize}
\end{minipage}
\end{table}

(9) The investigation of shallow-contact binary stars with
high mass ratios is important for the modern astrophysics because
they are considered as newly formed contact configurations, at the
beginning of contact evolution (\citealt{Qian+etal+2014}). The most
detailed studies of this type refer to the binaries AD Cnc (\citealt{Qian+etal+2007}), BI Vul (\citealt{Qian+etal+2013}) and CSTAR 038663 (\citealt{Qian+etal+2014}). They revealed that these cool, short-period (0.25--0.28
d), shallow-contact binaries exhibit strong magnetic activity
(including optical 0.2 mag flares of CSTAR 038663) and multiple
period changes. Recently we observed and modelled (in the same
way) shallow-contact W UMa's of H subtype (Table 8). On the
diagram fillout factor--mass ratio (Fig. 4) the targets from Table
8 fall into the bottom right (red triangles) due to their small
fillout factors (0.0-0.25) and high mass ratios (0.7--1.0). On the
same diagram the contact binaries with decreasing periods from the
sample of \cite{Yang+etal+2013} constitute cluster (black circles) to
the upper-left from our sample because they have intermediate
fillout factors (0.05--0.30) and moderate mass ratios (0.3--0.6).
One could guess that  the deep-contact W UMa's would form a third
cluster more left and upwards from the first two clusters on the
diagram. So, the diagram fillout factor--mass ratio acquires
evolutional meaning: through the common envelope phase the
position of a given star will describe a trace starting from the
bottom right and ending to the upper left side of the diagram.

More investigations of shallow-contact binary stars with
high mass ratios will provide more statistics of their global
parameters and opportunity for further study of the rapid
evolution of binary stars reached the contact stage. The presented
study is only a step in that direction.

\normalem
\begin{acknowledgements}
This work was supported partly by grants H08/21 and M08/02 of the Fund for
Scientific Research of the Bulgarian Ministry of Education and Science.
This publication makes
use of data products from the Two Micron All Sky Survey, which is
a joint project of the University of Massachusetts and the
Infrared Processing and Analysis Center/California Institute of
Technology, funded by the National Aeronautics and Space
Administration and the National Science Foundation. This research
also has made use of the SIMBAD database, operated at CDS,
Strasbourg, France, NASA's Astrophysics Data System Abstract
Service, the USNOFS Image and Catalogue Archive operated by the
United States Naval Observatory, Flagstaff Station
(http://www.nofs.navy.mil/data/fchpix/) and the photometric
software VPHOT operated by the AAVSO, Cambridge, Massachusetts.
The research was supported partly by funds of project RD 02-81 of
the Shumen University.
The authors are grateful to the anonymous referee for the
valuable notes and recommendations.
\end{acknowledgements}

\bibliographystyle{raa}
\bibliography{bibtex}

\end{document}